\newcommand*{\precise}{PRECISE}

\newcommand*{\hllm}{HLLM}
\newcommand*{\hstu}{HSTU}
\newcommand*{\learn}{LEARN}
\newcommand*{\sas}{SAS-Rec}

\newcommand*{\wechat}{WeChat}

\newcommand*{\ie}{\textit{i.e.}\@\xspace}
\newcommand*{\etal}{\textit{et. al.}\@\xspace}

\documentclass[conference]{IEEEtran}

\IEEEoverridecommandlockouts
\usepackage{cite}
\usepackage{amsmath,amssymb,amsfonts}
\usepackage{makecell}

\usepackage[linesnumbered,ruled,vlined]{algorithm2e}

\usepackage{xcolor} 

\SetAlgorithmName{Algorithm}{Algorithm}{Algorithm}
\IncMargin{0.5em}
\SetCommentSty{textnormal}
\SetNlSty{}{}{:}
\SetAlgoNlRelativeSize{0}
\SetKwInput{KwGlobal}{Global}
\SetKwInput{KwPrecondition}{Precondition}
\SetKwProg{Proc}{Procedure}{:}{}
\SetKwProg{Func}{Function}{:}{}
\SetKw{And}{and}
\SetKw{Or}{or}
\SetKw{To}{to}
\SetKw{DownTo}{downto}
\SetKw{Break}{break}
\SetKw{Continue}{continue}
\SetKw{SuchThat}{\textit{s.t.}}
\SetKw{WithRespectTo}{\textit{wrt}}
\SetKw{Iff}{\textit{iff.}}
\SetKw{MaxOf}{\textit{max of}}
\SetKw{MinOf}{\textit{min of}}
\SetKwBlock{Match}{match}{}{}
\usepackage{graphicx}
\usepackage{textcomp}
\usepackage{xcolor}
\usepackage{amsmath}
\usepackage{subfig}
\usepackage{multirow}
\usepackage{booktabs}
\def\BibTeX{{\rm B\kern-.05em{\sc i\kern-.025em b}\kern-.08em
    T\kern-.1667em\lower.7ex\hbox{E}\kern-.125emX}}
\begin{document}

\title{ PRECISE: \underline{P}re-training Sequential \underline{Re}commenders with \underline{C}ollaborat\underline{i}ve and \underline{Se}mantic Information
}

\newcommand{\AuthNameSep}{\hspace*{1em}}  
\newcommand{\AuthNameVSep}{\vspace{.5ex}}  %
\newcommand{\AuthNameVSpace}{\vspace{1.2ex}}
\newcommand{\AuthAffSep}{\hspace{1.5em}}
\newcommand{\AuthAffVSep}{}  
\newcommand{\AuthAffVSpace}{\vspace{.8ex}}
\newcommand{\AuthEmailSep}{\hspace{2em}}
\newcommand{\AuthEmailVSep}{\vspace{.2ex}}  %

\newcommand*{\AuthName}[2]{#2$^{#1}$}
\newcommand*{\AuthNameWithNote}[3]{\AuthName{#1}{#2}\titlenote{#3}}
\newcommand*{\AuthAff}[2]{$^{#1}${#2}}
\newcommand*{\AuthEmail}[2]{$^{#1}${#2}}
\newcommand*{\AuthEmails}[3]{$^{#1}${\{#3\}@#2}}

\author{
  \IEEEauthorblockN{Chonggang Song$^{\dag}$, Chunxu Shen$^{\dag}$\thanks{$^{\dag}$ Authors contributes equally to this paper.}, Hao Gu, Yaoming Wu, Lingling Yi$^*$\thanks{$^*$ Corresponding author.}, Jie Wen, Chuan Chen}
  \AuthNameVSpace
  \IEEEauthorblockA{
  Tencent Inc.
  }
  \AuthAffVSpace
  \IEEEauthorblockA{
    \AuthEmails{}{tencent.com}{jerrycgsong, lineshen, nickgu, aaronymwu,  chrisyi,  welkinwen, chuanchen}
  }
}

\maketitle

\begin{abstract}
 Real-world recommendation systems commonly offer diverse content scenarios for users to interact with. Considering the enormous number of users in industrial platforms, it is infeasible to utilize a single unified recommendation model to meet the requirements of all scenarios. Usually, separate recommendation pipelines are established for each distinct scenario. This practice leads to challenges in comprehensively grasping users' interests. Recent research endeavors have been made to tackle this problem by pre-training models to encapsulate the overall interests of users. Traditional pre-trained recommendation models mainly capture user interests by leveraging collaborative signals. Nevertheless, a prevalent drawback of these systems is their incapacity to handle long-tail items and cold-start scenarios.
With the recent advent of large language models (LLMs), there has been a significant increase in research efforts focused on exploiting LLMs to extract semantic information for users and items. However, text-based recommendations highly rely on elaborate feature engineering and frequently fail to capture collaborative similarities.


To overcome these limitations, we propose a novel pre-training framework for sequential recommendation, termed \precise. This framework combines collaborative signals with semantic information. Moreover, PRECISE employs a learning framework that initially models users' comprehensive interests across all recommendation scenarios and subsequently concentrates on the specific interests of target-scene behaviors. We demonstrate that 
\precise~precisely captures the entire range of user interests and effectively transfers them to the target interests. Additionally, we introduce practical training strategies that enhance the model's performance in real-world applications. Empirical findings reveal that the \precise~framework attains outstanding performance on both public and industrial datasets. \precise~has been deployed in multiple online recommendation scenarios within WeChat, and online A/B tests demonstrate substantial improvements in core business metrics.

\end{abstract}




\section{Introduction}
\label{Sec:introduction}

\begin{figure}[h]
\centering
\includegraphics[width=0.98\columnwidth]{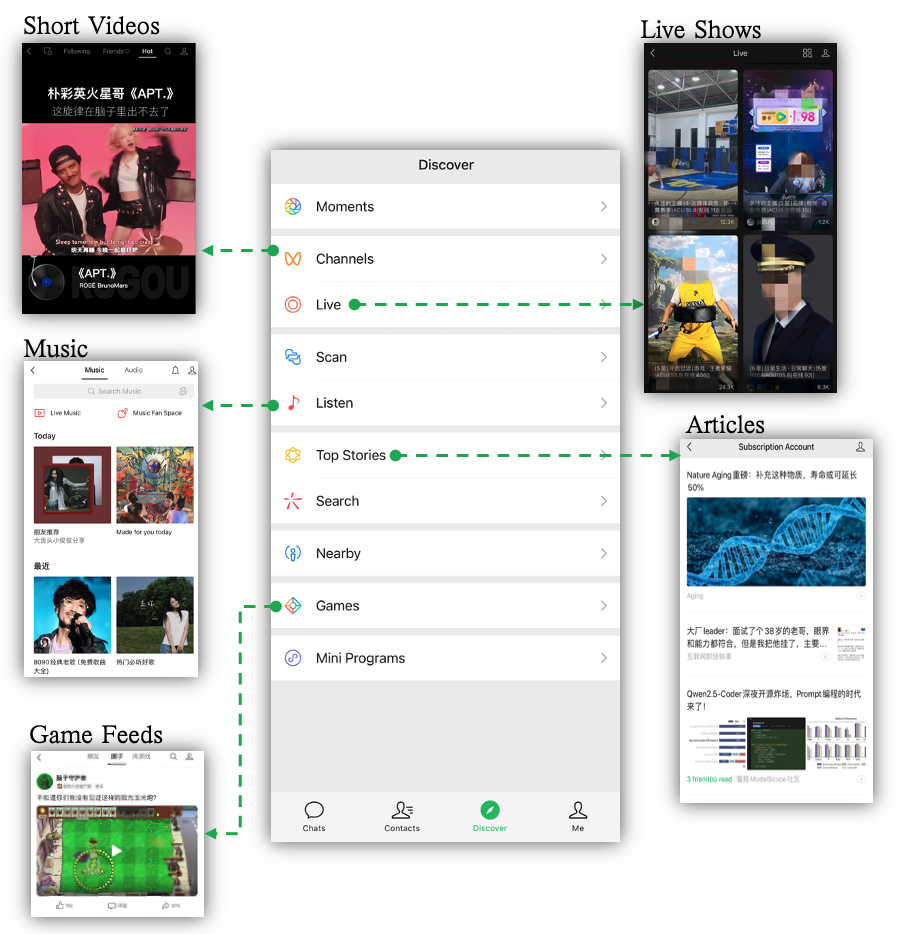}
\caption{WeChat Discover Tab. \textit{Channels}, \textit{Live}, \textit{Listen}, \textit{Top Stories} and \textit{Games} are five  recommendation scenarios with content of distinct modalities. \small(Real user images are blurred.)
}
\label{fig:intro-wechat}
\vspace{-0.58cm}
\end{figure}

In online recommendation systems,  diverse recommendation scenarios are often encountered. For instance, within the \wechat~\footnote{\wechat~is the largest online social platform in China with billion-scale active users. More in https://www.wechat.com/en/. } ecosystem, users can consume multiple types of content as illustrated in Figure~\ref{fig:intro-wechat}. Users can view short videos in \textit{Channels}, watch live shows in \textit{Live}, enjoy music in \textit{Listen}, read articles in \textit{Top Stories} and engage with games in \textit{Games}. 
Capturing user behaviors across the entire platform could provide a holistic  profiling of users' interests. 
However, due to the massive volume of user data, industrial recommendation systems seldom model multiple scenarios simultaneously and typically focus on providing recommendation services in a single scenario.

To incorporate behaviors across scenarios, many researchers have attempted to represent user behaviors in multiple scenarios with pre-training techniques~\cite{whiten2024,learn2024,hstu2024,ppm2024,sum2024,pinnerformer2022,pingcn2018,map2023}. 
Traditional sequential recommendation models predominantly utilize item IDs as the sole representation of items, capturing collaborative information between items through sequences of user interactions~\cite{lightgcn2020,hybridgnn2022,instantgnn2023,recgpt2024,idpretrain2024}. These approaches tend to bring embeddings of items that have been interacted with by similar users closer. 
Nevertheless, a significant limitation of these techniques lies in their inadequate representation of long-tail and cold-start items. This inadequacy leads to sub-optimal performance in recommending long-tail items to users who may exhibit interest, thereby obstructing the effective promotion of newly published items

With the advent of ChatGPT~\cite{chatgpt2023}, large language models (LLMs) have garnered widespread attention and applications across various domains. Recent efforts to integrate large language models with recommendation systems have yielded numerous successful experiences~\cite{whiten2024,hllm2024,learn2024,lcrec2024}. Current research in this domain primarily manifests in three forms:
\textit{Approach \uppercase\expandafter{\romannumeral1}} involves transforming the recommendation task into a dialog-style task, where recommendations are solicited from the large model through natural language descriptions, and the recommendations are directly obtained from the model's responses~\cite{llara2024,p52022}. While this method effectively leverages the model's world knowledge, it is challenging to deploy and launch in large-scale industrial recommendation systems.
 \textit{Approach \uppercase\expandafter{\romannumeral2}} extracts the intermediate representations of large models as supplementary information for users or items, aligning the representation space for recommendation tasks through adapter networks~\cite{learn2024,hllm2024,trawl2024,open2024}. Existing studies focus on  utilizing LLMs for representing the textual information, neglecting the ID representations that capture collaborative signals. These approaches may struggle to achieve satisfactory results when textual features of items are lacking or when feature engineering is not sophisticated.
\textit{Approach \uppercase\expandafter{\romannumeral3}} emulates the architecture and training methodologies of large models, specifically ChatGPT. It achieves this by constructing sequences of item IDs to mimic natural language token sequences~\cite{recgpt2024}. Subsequently, a causal decoder is utilized to predict the subsequent item. This kind of solution primarily varies from the traditional sequential recommendation pre-training strategies in terms of the training loss. However, it continues to encounter difficulties in dealing with long-tail items and cold-start scenarios.


In this work, we propose a sequential recommendation pre-training framework \precise~(\underline{P}re-training Sequential \underline{Re}commenders with \underline{C}ollaborat\underline{i}ve and \underline{Se}mantic Information) that leverages both item IDs and textual information for  the item representation. \precise~consists of three modules. In the first \textit{Embedding Fusion} module, we  concatenate ID representations and semantic representations generated by an LLM  to form a unified item representation. We introduce an expert network based on the Mixture of Experts (MoE) structure to balance the importance of token embeddings. In the second module named \textit{Universal Training}, we follow the model structure of ChatGPT and employ a decoder-only Transformer structure for predicting the next item in all-scene behavior sequences. 
This gives us a comprehensive understanding of user interest across multiple scenarios. 
In the last \textit{Targeted Training} module, we design task-specific loss functions to transfer holistic interests to  target tasks in the specific scenario. 
Note that \precise~framework is a combination of \textit{Approach \uppercase\expandafter{\romannumeral2}} and \textit{Approach \uppercase\expandafter{\romannumeral3}} of existing studies since \precise~takes in intermediate representations of LLM as an item feature and simulates the model structure of ChatGPT in the subsequent training steps.

Extensive online experiments have demonstrated that our proposed pre-training framework is capable of effectively capturing users' comprehensive interests across various scenarios, while rapidly adapting to the corresponding recommendation tasks through task-specific training.  By integrating both ID and textual representations, our model has achieved superior performance compared with state-of-the-art methods.  We also demonstrate that our framework is scalable in terms of data volume and the size of model parameters. Large-scale online A/B tests have validated the effectiveness of \precise~framework in real-world recommender platforms.

The contributions of this work can be summarized as follows:
\begin{itemize}
    \item We propose a pre-training framework \precise~for sequential recommendation in industrial platforms.  
    \precise~captures users'  full-domain knowledge in Universal Training and effectively transfers to specific tasks in specific domains with Targeted Training. 
    \item \precise~adopts LLM model for generating sophisticated semantic embedding for items  and 
    combines them with  collaborative information for item representation. We adopt an MoE-based Embedding Fusion module that selectively joins token embeddings to form semantic embeddings. 
    \item We introduce practical experience for training and deploying \precise~in real  online recommender systems. We conduct extensive experiments to validate the scalability and effectiveness of \precise~with both offline datasets and online experiments.
\end{itemize}

The rest of the paper is organized as follows. Section~\ref{Sec:relatedwork} introduces recent research on recommendation pre-training and  using LLMs for recommendation. Section~\ref{Sec:methodology} presents our proposed \precise~framework in detail and discusses practical experience for training \precise. In Section~\ref{sec:experiments-offline} and Section~\ref{sec:experiments-online}, we present offline experimental results and introduce serving schemes for deploying \precise~online. We conclude and discuss future work in section~\ref{Sec:conclusion}.

\section{Related Work}
\label{Sec:relatedwork}
In this section, we introduce  studies on pre-training recommendation models as well as recent works that adopt LLMs for recommendation.
\subsection{Pre-training in Recommendation}
\label{Sec:relatedwork-seq}
Recently, pre-trained models have achieved significant success in the field of Natural Language Processing (NLP). The predominant approach involves pre-training models on large-scale datasets followed by fine-tuning on the target domain datasets, such as BERT~\cite{bert2019} and LEGAL-BERT~\cite{legalbert2020}. Research in recommendation systems has also sought to harness the power of the pre-training paradigm  to capture users' long-term and cross-scene interests. 

Pre-training recommendation models  often unfold in two distinct manners. 
The first type of approach utilizes  multi-modal information for pre-training~\cite{unisrec2022,map2023,learn2024,multimodal2024,ppm2024,word2023,multimodal2023,missrec2023}.
Hou~\etal~\cite{unisrec2022} and Yang~\etal~\cite{word2023} construct text-based attributes to represent item features and use behavioral sequences to align semantic representation with behavioral interest. Li~\etal~\cite{multimodal2024} and Zhang~\etal~\cite{multimodal2024} construct advanced structures for fusing and encoding multi-modal information  for interest matching in sequence recommendation. 
Wang~\etal~\cite{missrec2023}  design an encoder-decoder model where the encoder captures multi-modal features and the decoder aligns feature embedding with user interest. 
Gao~\etal~\cite{ppm2024} and Lin~\etal~\cite{map2023} pre-train feature encoding module based on modality information and plug pre-trained modules into ranking models to enhance feature embedding.
These approaches lack the utilization of ID representation, resulting in suboptimal  performance when multi-modal features are not adequate.

The second type of approach involves pre-training on user or item IDs. Due to reliance on ID representation, the majority of such studies pre-train models on target-domain datasets~\cite{instantgnn2023,hybridgnn2022,ssgnn2021,sslrec2021,hstu2024}.  
Kang~\etal~\cite{sasrec2018} develop a self-attention-based sequential model (\sas) that has been widely adopted by many sequential recommenders~\cite{sasrec2023}. 
Works in~\cite{hybridgnn2022,ssgnn2021,instantgnn2023}  employ graph structures to describe user-item interactions and adopt contrastive learning for generating representations for specific user/item IDs. 
Yao~\etal~\cite{sslrec2021} adopt a multi-task self-supervised learning framework for large-scale items.
Recent work by Zhai~\etal~\cite{hstu2024} proposes a generative recommender that adopts an accelerated self-attention mechanism for modeling behavior sequences. 
Those solutions do not consider the incompatible ID  transition problem from the pre-training model to downstream tasks since pre-training and downstream tasks focus on the same scope of recommendation scenarios.
However, our platform inherently offers a variety of recommendation scenarios, necessitating the integration of multi-scenario behaviors into user representation. Concurrently, we design a module for tuning model parameters to target scenarios, enabling the model to rapidly adapt to downstream tasks. Note that 
our framework naturally involves learning from a full scenario before focusing on a specific sub-scenario, thus eliminating the issue of ID migration.
Recent efforts~\cite{idpretrain2024} have been directed towards cross-domain ID mapping in recommendation systems.  Authors build a cross-domain ID-matcher learned by  behavioral and modality information that allows IDs to appear in the source domain can transfer to IDs in target domains.
This work falls within the scope of cross-domain recommendation. It is not addressed by our current pre-training framework since our work focuses on the transition from general domains to specific sub-domains.



\begin{figure*}[h]
\centering
\includegraphics[width=1.98\columnwidth]{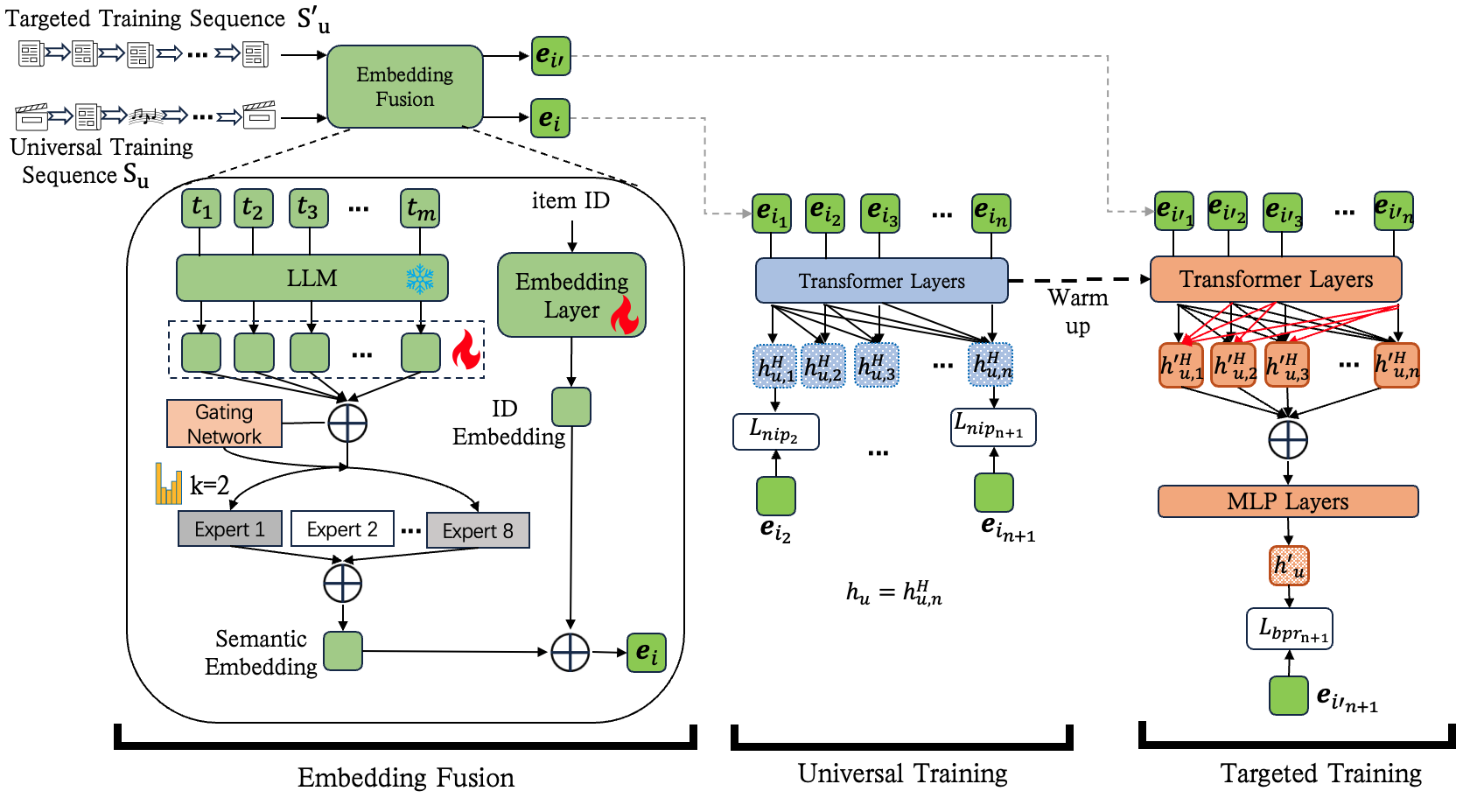}
\caption{The overview of \precise~framework. \small(\precise~consists of three modules: Embedding Fusion, Universal Training and Targeted Training. Figure best viewed in color.) }
\label{fig:method-framework}
\vspace{-0.3cm}
\end{figure*}

\subsection{LLMs for Recommendation}
\label{Sec:relatedwork-llm}
With the huge success of LLMs, many researchers have endeavored to integrate the capabilities of LLMs into their recommendation models. 
The first type of study involves transforming user behaviors into natural language descriptions and employing prompt-tuning to enable the LLMs to directly infer subsequent recommendation outcomes~\cite{llara2024,userllm2024}. Liao~\etal\cite{llara2024} employ a projector to align ID embeddings with the LLM's input space and adopt a curriculum learning strategy to progressively guide the learning of LLM.   Ning~\etal~\cite{userllm2024} integrate pre-trained user embeddings with LLMs with cross-attention, enabling LLMs to generate responses based on the context of a user's past behaviors.
These methodologies necessitate the deployment of the LLM model to infer online user requests, thereby presenting significant challenges in fulfilling latency requirements within industrial systems. 
The second type of research efforts has focused on extracting semantic information through LLMs, incorporating this semantic data as features into subsequent recommendation tasks~\cite{learn2024,hllm2024}. Jia~\etal~\cite{learn2024} adopt LLMs to generate item embeddings and adjust a twin-tower structure for user and item for interest matching. Chen~\etal~\cite{hllm2024} employs a two-tier model for item and user respectively. Both tiers are initiated with LLM parameters and trained with recommendation tasks.
A challenge inherent in this method is the difficulty in capturing collaborative signals, rendering it less effective in scenarios where textual information is scarce. For example, in scenarios with  short videos or images, it is challenging to obtain reliable feature representations.
The third approach draws inspiration from the training paradigm of models like ChatGPT~\cite{recgpt2024}, utilizing behavioral data from recommendation scenarios to construct training corpora. These methods leverage the structure and training techniques of LLMs to create a dataset that reflects the actions and interactions within the recommendation context. Work in~\cite{recgpt2024} further designs a prompting mechanism to fine-tune models for predicting future items. 
Our proposed approach is a combination of the second and third approach. To the best of our knowledge, this work is a pioneering study that combines ID embedding with representations generated by LLMs in industrial recommendation systems.

\begin{table}[t]
\centering
\caption{Frequently Used Notations.}
\vspace{-0.1cm}
\centering
\small
\begin{tabular}{ll}
\toprule
Symbols & Descriptions \\
\hline
$U$, $I$ & The sets of users and items respectively \\
$u$, $i$, $t$ & User ID, item ID and token ID respectively \\
$S_u$ & The behavior sequence of  $u$ in all scenes \\
 $S'_u$ & The behavior sequence of  $u$ in the target scene \\
$T_i$ & The token sequence of item $i$ \\
$\boldsymbol{e}_i,E_u$ & Embedding of item $i$ \& Embedding matrix of $u$\\
$K$, $k$ & The total number and  chosen number of MoE experts \\
$H$ & Total number of Transformer layers \\
$h_{u,i}^l$ & Item $i$'s embedding output from the $l^{\text{th}}$ Transformer\\ & layer  where $i\in S_u$ \\
$L_{nip}$ & The Next-item Prediction loss \\
$L_{bpr}$ & The Bayesian Personalized Ranking loss \\
$f_{UT}$, $\theta_{UT}$ & Transformer model and parameters in Universal Training\\
$f_{TT}$, $\theta_{TT}$ & Transformer model and parameters in Targeted Training\\
\bottomrule
\end{tabular}
\vspace{-0.35cm}
\label{tab:notations}
\end{table}

\section{Preliminary}
\label{Sec:preliminary}
Assume we have a set of users $u \in U$ and a set of items $i\in I$, $u$ and $i$ represent user and item IDs respectively. For each user $u$, $S_u = [ i_{u,1}, i_{u,2}, \cdots ,i_{u,n}] $  denotes the sequence of historical interacted items in chronological order, where $n=|S_u|$ indicates the number of interacted items. 
For each item $i$, $T_i = [t_{i,1}, t_{i,2}, \cdots, t_{i,m}]$ denotes the text tokens of item $i$ in its description, where $m=|T_i|$ is the length of items' maximum token length. 
For  article $i$, $T_i$ can be extracted from article titles and texts. For a short video $i$, $T_i$ can be retrieved from video titles and hashtags. For live show $i$, $T_i$  represents show titles and texts recognized from speech. 

Given a user $u$'s behavior sequence $S_u$ of length $n$, the task of sequential recommendation is to predict the next item  $i_{u,n+1}$ that user $u$ is most likely to interact with. The problem can be formulated as 
\begin{equation}
    \label{Eq:problem}
    \mathop{\arg\max}\limits_{\hat{i}\in I}P( i_{u,n+1} = \hat{i}|S_u)
\end{equation}
where $\hat{i}$ is the predicted item ID given user sequence $S_u$. The notations used in this paper are summarized in Table~\ref{tab:notations}.

\section{Methodology}
\label{Sec:methodology}
In this section, we first  introduce  the overall framework of \precise~and then elaborate on each module in  detail. Lastly, we discuss practical training strategies for achieving optimal results in real-world recommendation systems.

\subsection{Overall Framework}
\label{Subsec:methodology-framework}
In this paper, we propose a pre-training  framework for sequential recommendation named \precise. The overall framework is depicted in Figure~\ref{fig:method-framework}. As shown in the figure, \precise~consists of three modules. In the \textit{Embedding Fusion} module, we generate item embeddings for each item in a user's behavior sequence. We employ an embedding module to obtain the ID embeddings, concurrently utilizing a Large Language Model (LLM) to generate token representation to initialize  token embeddings. Note that both ID embedding and token are trainable variables in our framework. Subsequently, a Mixture of Experts (MoE) network is applied to distill the token embeddings into semantic representation, which is further concatenated with ID embedding to form the final item embedding. 
In the \textit{Universal Training} module, we use users' all-scene behavior sequences to train a Transformer model. Item embeddings from the Embedding Fusion module are fed into the Transformer model, where we employ a Next-item Prediction (NIP) loss for each item within the sequence. 
In the last \textit{Targeted Training} module, we use users' target-scene behavior sequences to train a Transformer model. 
For instance, during Universal Training, we utilize a variety of user behaviors across all scenarios on the recommendation platform, such as watching short videos and reading articles. However, in the Targeted Training phase, we focus on a single task, such as \textit{clicking} on \textit{articles}. Therefore, the data used in this Targeted Training consists of sequences of specific user behaviors within a single scenario. The user behaviors are chosen based on the requirements of our downstream tasks. 
As shown in Figure~\ref{fig:method-framework}, we warm-start the Targeted Training model with parameters trained in  Universal Training and further train the model with target-scene behavior sequences. In this step, we adopt a Bayesian Personalized Ranking (BPR) loss, which is more aligned with the recommendation task.  In the following, we detail the specifics of each module.

\subsection{Embedding Fusion Module}
\label{Sebsec:methodology-fusion}
In the context of sequential recommendation, the representation of sequences as a series of embeddings allows for the modeling of complex user-item interactions. The Embedding Fusion module takes in the interacted items of user $u$, which are placed in chronological order. It constructs the embedding matrix $E_u$ where each row corresponds to the embedding $\boldsymbol{e}_i$ of an item $i$, as shown in Figure~\ref{fig:method-framework}.

Each item contains two features: item ID and item text. In  \wechat, items of various modalities can be represented in the form of texts. Therefore, in this paper, we primarily utilize LLMs for semantic embedding. Note that LLMs can be replaced by multi-modal embedding methods when available.
In the Embedding Fusion module, we first perform  an  embedding lookup of the item ID to obtain the ID embedding.
We then obtain  semantic embeddings by inputting the textual tokens of an item to the  LLM. The token embeddings are  arranged in the order of the original text.  
The token embeddings of item $i$ are represented as  $\bold{x}_i$. Formally, we have 
\begin{equation}
    \label{eq:xi}
    \boldsymbol{x}_i =LLM(T_i)
\end{equation}
where $LLM$ represents invoking an LLM model for token encoding.

We then feed LLM embeddings to an MoE module. We have $K$ expert networks that are essentially attention modules, denoted as $[Attn_1, Attn_2, \cdots, Attn_K]$ .  Each attention module computes a weighted sum of  token embeddings, aggregating $m$ token  embeddings into one embedding of dimension $d$.
Subsequently, a gating network $W^{gate}$ is employed to compute the score of each expert. Based on the scores, we select the top $k$ experts to activate. Then, the $k$ experts'  weights are computed with a $softmax$ function.
The computation process of the gating network can be formalized as follows. 
\begin{equation}
    \label{eq:gate}
    gate(\boldsymbol{x}_i) =softmax(topk(\boldsymbol{x}_i \cdot W^{gate}))
\end{equation}

Subsequently,  chosen experts' output embeddings are pooled together as the item's semantic embedding while unchosen experts' weights are set to $0$.  The semantic embedding is concatenated with ID embedding to form the item's final representation. Formally, we have 
\begin{equation}
\label{eq:itemembedding}
\boldsymbol{e}_i = ID(i)\oplus MoE(\boldsymbol{x}_i) = ID(i)\oplus \sum\limits_{j=1}^K (gate(\boldsymbol{x}_i)_j\cdot Attn_j(\boldsymbol{x}_i) )
\end{equation}
where  $ID$ represents the lookup function of ID embedding, $gate(\boldsymbol{x}_i)_j$ is the weight for expert $j$ determined by  the gating network and  $Attn_j$ is the attention function for expert $j$. The embeddings of items in user $u$'s behavior sequence are stacked together as $E_u$, and is further fed to Universal Training and Targeted Training modules.

\subsection{Universal Training Module}
\label{Sebsec:methodology-pre}
In the  Universal Training module,  we  train a decoder-only Transformer model to represent the all-scene behavior sequences.
Transformer architecture has emerged as a powerful model design in the domain of sequential recommendation~\cite{attn2017,joint2024}. 
Each Transformer layer is composed of a Masked Multi-head Self-attention module (MMS) and a Feed-forward Network (FFN). The implementation details of the Transformer module can be found in~\cite{attn2017}. 
Masked Multi-head Self-attention module employs a masking matrix for modeling  item correlations in sequences. The lower
triangular part of the masking matrix is set to $0$ and the remaining part $-\infty$, allowing each item to attend to past items, but preventing them from attending to future items.  Feed-forward Network outputs a list of embeddings, where the embedding at  position $i$ predicts the corresponding $i+1^{\text{th}}$ item in the sequence. We present the  computation of
Universal Training as follows
$$h_u^0 = E_u $$
$$h_u^l = TransformerBlock(h_u^{l-1})$$
$$h_u=h_{u,n}^H = f_{UT}(E_u | \theta_{UT} )[n]   $$
where $E_u$ is the embedding matrix of user $u$'s behavior sequence, $l = [1, \cdots, H]$ represents the $l^{\text{th}}$ layer in the Transformer model, $h_u$ represents the final representation of user $u$  and it is given by the embedding at position $n$ in the last Transformer layer's output as shown in Figure~\ref{fig:method-framework}.  $f_{UT}$ represents the Transformer model with $\theta_{UT}$ being all trainable parameters in the Universal Training phase.

Given the above model, we then introduce the loss function used in the Universal Training. The model is expected to estimate the next interacting item for each position in the sequence. In language models, a $softmax$ function on all tokens is utilized for picking the predicted word token. However, in recommendation tasks, the candidate pool contains hundreds of millions of items, rendering it impossible to construct a full $softmax$ matrix of all candidate items. To overcome this problem, for each item in a sequence $S_u$, we sample items from other sequences as negative samples to form a negative sample set $N_u$ and use cross-entropy loss to guide the model training. Formally, we have 

\begin{multline}
\label{eq:nip}
L_{nip} =  
-\sum\limits_{u\in U,i\in S_u}\frac{ log(h_{u,i}^H \cdot 
 \boldsymbol{e}_{i+1}) }{  log(h_{u,i}^H \cdot 
 \boldsymbol{e}_{i+1}) + \sum\limits_{j\in N_u, j\neq i+1}log(h_{u,i}^H \cdot 
 \boldsymbol{e}_{j}) }
 \end{multline}
where $N_u$ is a set of items randomly sampled from other sequences, $h_{u,i}^H$ gives the  embedding of item $i\in S_u$ in the final Transformer block and $\boldsymbol{e}_{u,i+1}$ is the item embedding of the next item in $S_u$.   We aim to maximize the dot product of the computed embedding of item $i$ and the item representation of item $i+1$ in each user $u$'s behavior sequence $S_u$. 

\subsection{Targeted Training Module}
\label{Sebsec:methodology-finetune}
\begin{figure}[h]
\centering
\includegraphics[width=0.9\columnwidth]{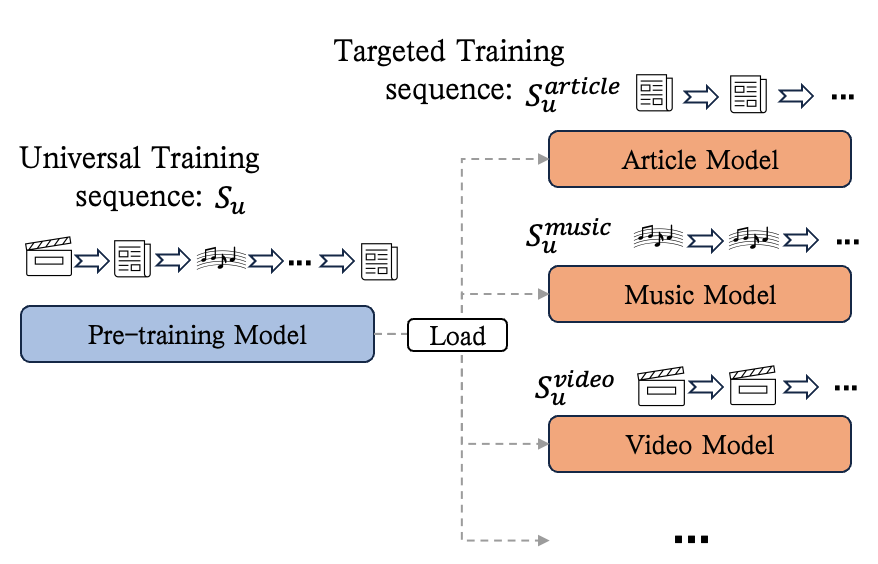}
\caption{Universal and Targeted Training Scheme of \precise. }
\label{fig:method-load}
\end{figure}

In the Targeted Training module,  we  use  target-scene behavior sequences to train a Transformer model $f_{TT}$. 
Firstly, we  warm-start all parameters in $f_{TT}$ from the Transformer model $f_{UT}$ in  Universal Training module. Subsequently, the model parameters are trained from behavioral data specific to particular scenarios or tasks. Since Universal Training utilizes behaviors from all scenarios, the  parameters from  Universal Training can be adapted to many Targeted Training models where each model corresponds to a distinct scenario as illustrated in Figure~\ref{fig:method-load}.

To better accommodate the recommendation task, we have made two modifications in model training. Firstly, when computing the attention weights between items in behavior sequences, we have adopted bi-directional attention for all items~\cite{bert4rec2019}. This implies that within each Transformer block, no masking matrix is used. The advantage of this approach is that  the representation of each item in the sequence is computed based on the complete sequence.
Secondly, to utilize all item embeddings within each user's behavior sequence, we concatenate all item embeddings from the final Transformer layer and adopt a multilayer perceptron (MLP) network to compute the final embedding of users as shown in the Targeted Training part in  Figure~\ref{fig:method-framework}. Formally, we have 

$$h_u^{\prime 0} = E'_u $$
$$h_u^{\prime l} = TransformerBlock^{NoMask}(h_u^{\prime l-1})$$
$$h'_u = f_{TT}(E'_u|\theta_{TT}) =  MLP(CONCAT(h_{u,1}^{\prime H},\cdots, h_{u,n}^{\prime H}))$$
where $l = [1, \cdots, H]$ represents the $l^{\text{th}}$ layer in the Transformer model and $S'_u$ is the  behavior sequence of user $u$ in the target scene. $h'_u$ represents the final user embedding and it is given by an MLP network following the Transformer blocks.  $f_{TT}$ represents the  model structure with $\theta_{TT}$ being all trainable parameters in the Targeted Training phase. 

With this design, earlier items  in the sequence can 'see' items later in the sequence,  Next-item Prediction (NIP) loss becomes infeasible. In this phase, we employ a Bayesian Personalized Ranking (BPR) loss  that, given a sequence of length $n$, only predicts the $n+1^{\text{th}}$ item.   
BPR loss is commonly adopted in recommendation tasks as it directly optimizes the personalized ranking of items, which is a core objective in recommendation systems~\cite{bpr2009}. Formally, we have 
\begin{equation}
\label{eq:bpr}
L_{bpr} = - \sum\limits_{u\in U}\sum\limits_{u'\in U, u'\neq u} log(h'_{u} \cdot \boldsymbol{e}_{u,n+1} - h'_{u'} \cdot \boldsymbol{e}_{u,n+1} )
\end{equation}
where $h'_{u}$ is the user embedding computed from the  user sequence $S'_u$ with length $n$ and $\boldsymbol{e}_{u,n+1}$ is the item embedding of the next item following the given user sequence.

\subsection{Discussion on Training Strategies}
\label{Sebsec:methodology-discussion}
In this section, we elucidate the challenges and issues encountered during the training process of models along with our solutions.

\subsubsection{Periodical Warm-up}
\begin{figure}[t]
	\begin{center}
		\hspace*{-0.1in}
		\subfloat[\small Rate of Items Untrained in the Universal Training w./w.o. Periodical Warm-up in Targeted Training.]{
			\includegraphics[scale=.375]{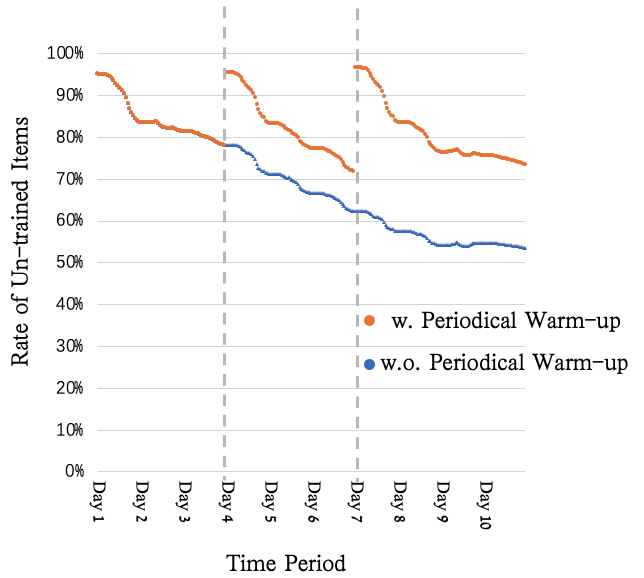}
			\label{fig:problem-period}
		}
	\hspace*{0.045in}	
		\subfloat[\small Cosine Similarity Distribution of ID Embedding w./w.o. Alternate Training among Similar Items Pairs.]{
			\includegraphics[scale=.375]{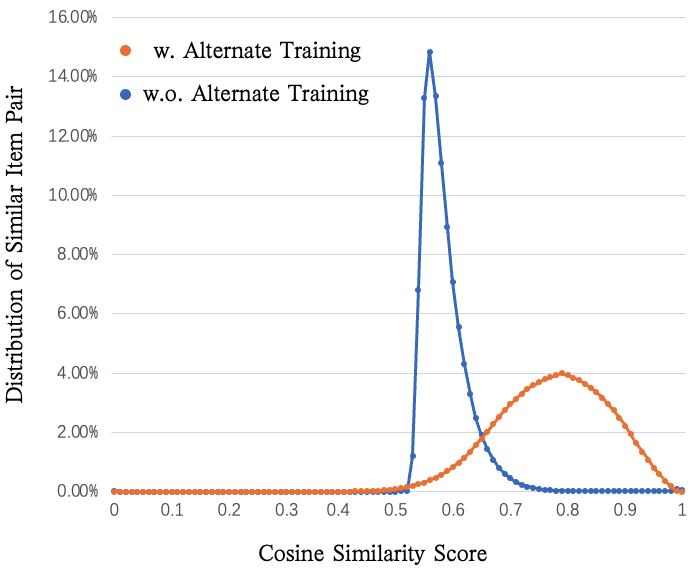}
			\label{fig:problem-alter}
		}		
		\caption{Illustration of Training Problems.}
		\label{fig:problem}
	\end{center}
    \vspace{-0.5cm}
\end{figure}

In large language models, the pre-training process is conducted on a vast corpus, while fine-tuning focuses on corpora from a specific domain. This approach is applicable when the total number of tokens remains constant. However, in recommendation scenarios, new items are produced every minute. If the Targeted Training module  warm-start the parameters from the Universal Training model only  at the starting point of model training,  items produced after the starting point  will not have been trained by the Universal Training module. 
We illustrate this issue with Figure~\ref{fig:problem-period}. Assume that we start Targeted Training on Day 1 and do not warm up the parameters with the  Universal Training model afterwards, the proportion of exposed items trained with all-scene behaviors decrease significantly  each day. However, if we periodically warm up the parameters with Universal Training model, the proportion of pre-trained items remains above $70\%$. As we show in experiments, Universal Training is critical in improving the recommendation performance and hence we periodically warm up the Targeted Training model with parameters from the Universal Training model.
The specific warm-up cycle can be determined based on the iteration rate of new items on specific recommendation platforms.

\subsubsection{Alternate Training}
One important innovation of this work is the combination of item ID representations with item text representations. During model training, we observe that since text representations are derived from well-trained large language models, the model training heavily relies on semantic embeddings, leading to insufficient training of ID representations. 
We illustrate this issue with Figure~\ref{fig:problem-alter}.
We sample item-pairs $<i,j>$ that are considered similar to each other given their overall item embeddings $<\boldsymbol{e}_i, \boldsymbol{e}_j>$. Afterwards, We  specifically compute the cosine similarity of $<i,j>$ with their ID embeddings. 
The blue line represents the similarity distribution of ID embeddings trained along with token embedding while the orange line represents the similarity distribution of ID embeddings trained with token embedding fixed.
As we can see in Figure~\ref{fig:problem-alter}, when ID embeddings are trained without token embeddings  held constant, the cosine similarity  remains approximately around $0.55$. This value is noteworthy as it is only slightly higher than the expected similarity of $0.5$ for randomly initialized ID embeddings, suggesting that  ID embeddings have undergone minor degree of update. Conversely, when token embeddings are held constant, the resultant ID embeddings exhibit a more distinguishable distribution. 

To address this issue, we adopt an alternate training strategy. We initially fix the token embeddings and only train  the ID embedding layers. When the ID representations have converged, we  allow the training of token embeddings.




\section{Offline Experiments}
\label{sec:experiments-offline}
In this section, we conduct experimental studies on offline datasets to verify \precise’s effectiveness and scalability, followed by ablation studies to validate the contribution of each component in \precise.


\subsection{Experimental Setup} 
\textbf{Datasets.}
We use two  datasets for offline experiments.
The first dataset is the Amazon book review dataset\footnote{https://www.kaggle.com/datasets/mohamedbakhet/amazon-books-reviews} which consists of users' reviews on purchased books. These reviews span from a low of 1 to a high of 5, with 1 representing minimal interest and 5 indicating considerable enthusiasm. We regard the reviews with scores of 3 or above classified as positive feedback. We adopt the 'leave-one-out' strategy for testing and only collect users' purchased sequence of no less than 3 items as our training examples. Notably, to ensure comparability with ID-centric approaches, our testing dataset includes only items that have been previously presented in the training dataset. Item representation is accomplished by incorporating the book title, description, and categories as textual elements. This dataset is named \textit{Amazon} and  its detailed statistics are given in Table~\ref{tab:dataset}.
\begin{table}[htp]
	\centering
     \renewcommand\arraystretch{1.2}
	\caption{Dataset Details} 
	\begin{tabular}{c|c|c|c|c}
		\toprule
		Dataset & Task &   \#User  & \#Item & \#Behavior \\
		\hline
		\multirow{1}{*}{Amazon}	&	All & 1,008,972 & 221,997 & 2,438,212    \\ 
		\hline
		\multirow{3}{*}{WeChat}	&	AllScene & 109 million & 140 million & 11 billion  \\ 
		& Click & 74 million & 37 million & 4 billion \\ 
		& Share &  33 million & 95 million & 550 million \\
		\bottomrule
	\end{tabular}
	\label{tab:dataset}
\end{table}

The  second dataset is  collected from the online article recommendation system on WeChat. We collect behaviors from randomly sampled $10\%$ users during the period from April  2023 to November 2024.  
For Universal Training, we collect users' clicked articles from all scenarios, \ie Top Stories, Recommendation Box and so on. We call this dataset \textit{WeChat-AllScene}, which is used for Universal Training in \precise.
We then consider two different tasks to test \precise's transferability in cross-scene and cross-task  situations. 
The first task is \textit{clicking} in the \textit{Recommendation Box}. For this task, we collect users' clicked articles only in the Recommendation Box scenario and order user behaviors by time. 
The second task is \textit{sharing} articles from \textit{all scenarios}. For this task, we collect users' shared articles in all scenarios  in chronological order. We call these datasets \textit{WeChat-Click} and \textit{WeChat-Share} respectively. We use article title and categories as article texts. 
Testing datasets consist of behaviors occurring on the day  after the end time of the training dataset.
Details are given in Table~\ref{tab:dataset}. 
Note that there are significant differences in user behaviors between the two targeted scenarios. Clicking is more concentrated, with $74$ million individuals performing a total of $4$ billion clicks, all of which are focused on $37$ million items. In contrast, sharing behavior is distributed across a variety of different items, with only $33$ million individuals engaging in sharing, yet the number of  shared  items amounts to  $95$ million. We use such distinct datasets to verify \precise~is adaptable to different target tasks.

\begin{table*}[htp]
	\centering
    \renewcommand\arraystretch{1.2}
    
	\caption{Performance of \precise~UT and SOTA methods. \small (Best performance in bold and runner-up performance underlined.)} 
	\begin{tabular}{c|c|c|c|c|c|c|c}
		\toprule
		Dataset & Method &   R@10  & R@30 & R@50 &  N@10  & N@30 & N@50 \\
		\hline
		\multirow{7}{*}{Amazon}	&	SAS-Rec & 0.0356 & 0.0662 & 0.0673  & 0.0177 & 0.0244 & 0.0249   \\ 
		& HSTU & 0.0387  & 0.0701 & 0.0722 & 0.0197 & 0.0259 & 0.0269  \\ 
		& 		LEARN & 0.0282 & 0.0500 & 0.0590 & 0.0164 & 0.0222 & 0.0229   \\  
		& 		HLLM & \underline{0.0419} & \underline{0.0792} & \underline{0.0846} & \underline{0.0227} & \underline{0.0277} & \underline{0.0282}  \\  
            &       \precise-UT & \textbf{0.0439} & \textbf{0.0888} & \textbf{0.0972} & \textbf{0.0269} & \textbf{0.0292}  & \textbf{0.0311}  \\  \cline{2-8}
            &       Improve & $+4.60\%$  & $+12.12\%$ & $+14.89\%$ & $+18.5\%$ & $+5.4\%$ & $+10.3\%$ \\
		\hline
		\multirow{7}{*}{WeChat-AllScene}	&	SAS-Rec & 0.0097 & 0.0199 & 0.0251 & 0.0072 & 0.0119 & 0.0138   \\ 
		& HSTU & 0.0159 & 0.0291 & 0.0409 & 0.0109 & 0.0122 & 0.0146  \\ 
		& 		LEARN & 0.0082 & 0.0154 & 0.0203 & 0.0058 & 0.0094 & 0.0107  \\  
		& 		HLLM & \underline{0.0174} & \underline{0.0332} & \underline{0.0434} & \underline{0.0125} & \underline{0.0144} & \underline{0.0162}  \\  
            &       \precise-UT & \textbf{0.0197} & \textbf{0.0359} & \textbf{0.0460} & \textbf{0.0139} & \textbf{0.0165} & \textbf{0.0185}  \\  \cline{2-8}
            &       Improve & $+13.22\%$ & $+8.13\%$  & $+6.00\%$  & $+11.2\%$  & $+14.58\%$  & $+14.20\%$  \\
		\bottomrule
	\end{tabular}
	\label{tab:overall}
\end{table*}
\textbf{Methods.}
We compare the following state-of-the-art pre-training methods that are feasible for large-scale recommendation systems:

\textit{\precise} is the proposed framework consisting of three major components, \ie Embedding Fusion, Universal Training and Targeted Training. When there is no targeted scenario, we  use embeddings trained from Universal Training for comparison. This model is termed \precise-UT in the following discussions.

\textit{SAS-Rec} is a sequential recommendation model that adopts a self-attention mechanism to determine relevant history items for predicting the current item in users' behavior sequences~\cite{sasrec2018}. We adopt the \sas~implementation in~\cite{sasrec2023} that  is competitive in sequential recommendation. Note that SAS-Rec focuses on utilizing collaborative information and no semantic information is used. 

\textit{HSTU} is a recent generative recommendation model proposed in~\cite{hstu2024}. It design a hierarchical sequential transduction  unit for sequence encoding to fit recommendation task. Note that HSTU also focuses on modeling ID embedding and no textual features are used.

\textit{LEARN} is an adaptive framework that transforms knowledge from LLM models to semantic embeddings that can be adapted in recommendation tasks~\cite{learn2024}. LEARN omits ID modeling  and merely adapts textual information for modeling collaborative behaviors.

\textit{HLLM} is a hierarchical large language model that  utilizes an LLM structure for item modeling, followed by another LLM structure for user modeling~\cite{hllm2024}. HLLM also focuses on semantic modeling only and omits ID modeling. Note that authors attempt to combine pre-trained ID embedding with HLLM embedding in the output layer and found a decrease in performance. We handle this challenge with our Embedding Fusion module in \precise.  

We adopt the following configuration for all models. All models are trained from scratch with at most $20$ epochs. Batch size is set to $128$ and $6$ layers of Transformer blocks are used. The embedding dimension is set to $128$ for both ID and token embedding and hidden layer dimension for  the  feed-forward network is set to $512$. We adopt the Qwen2\footnote{https://qwen2.org/} model for all models that require an LLM for generating token representations.

Note that  state-of-the-art solutions lack a training procedure specific to targeted scenarios. For fair comparisons, we first compare \precise-UT with  existing approaches on the academic dataset \textit{Amazon} and industrial dataset \textit{WeChat-AllScene}. Next, we perform ablation and scalability study on \precise-UT with the \textit{WeChat-AllScene} dataset. 
Subsequently, we validate the efficacy of the  entire \precise~framework  on the targeted tasks with \textit{WeChat-Click} and \textit{WeChat-Share} datasets.

\textbf{Metrics.} We evaluate our proposed method with  two commonly used metrics. Recall at top-k 
 (R@K)  is a metric used in recommendation systems and information retrieval to evaluate the proportion of actual relevant items that are included in the top-k recommended items. Normalized Discounted Cumulative Gain (N@K), is another metric used to assess the quality of the top-k recommended items. It takes into account both the position of relevant items in the recommendation list and their relevance scores.

\subsection{Overall Performance} 
Table~\ref{tab:overall} shows the  performance of \precise-UT compared with all SOTA methods in both academic and industrial datasets. As we can see, \precise-UT outperforms all baseline methods in both Amazon and \wechat-AllScene datasets. This is because \precise-UT combines both ID and text token embedding to enhance item representation while baseline approaches either utilize ID embedding or text embedding alone, leaving their combination unexplored. We also notice that the \hllm~model is a strong baseline model, indicating that items can be effectively represented by text. However, \learn~performs poorly, which is attributed to the fact that it employs fixed token embeddings, leaving the token embedding and LLM parameters untrained. However, both \hllm~and  \precise~permit updates to the token embeddings, thereby enabling more profound adaptation to parameters induced by collaborative signals.
In the context of pure ID models, \hstu~demonstrates a superior performance over \sas~due to its improvements in  attention mechanism to fit recommendation task. However, neither
\sas~nor \hstu~gives the best performance since they rely heavily on ID representation and perform poorly on cold items.

\begin{table}[htp]
	\centering
	\caption{Performance on Cold Items \small (Items  interacted less than $1000$ times in \wechat-AllScene are used for evaluation.)  } 
    \renewcommand\arraystretch{1.2}
	\begin{tabular}{c|c|c|c|c}
		\toprule
		 Method &   R@10  & R@30  &  N@10  & N@30  \\
		\hline
		\sas  & 0.0069 & 0.0127  & 0.0041 & 0.0056   \\ 
		 HSTU & 0.0079 & 0.0138 & 0.0049 & 0.0064  \\ 
  		LEARN & 0.0078 & 0.0246 & 0.0068 & 0.0143   \\  
        HLLM & \underline{0.0227} & \underline{0.0462} & \underline{0.0131} & \underline{0.0192}   \\  
        \precise-UT & \textbf{0.0260} & \textbf{0.0526} & \textbf{0.0148} & \textbf{0.0218}   \\
    \hline
    Improve & +14.5\% & +13.85\% & +12.97\% & +13.54\%   \\
		\bottomrule
	\end{tabular}
	\label{tab:colditem}
    \vspace{-0.5cm}
\end{table}

Next, we evaluate model performances on cold or long-tail items. 
Sequence recommendation models based on ID modeling often exhibit superior performance for popular items but are less accurate for long-tail item recommendations. In \precise, we integrate semantic representation generated by LLM to enhance model performance for cold items. We rank all items by the frequency of user interactions, from highest to lowest, and then designate the top $20\%$ of items with the highest popularity as hot items and the remaining $80\%$ with lower popularity as cold items~\cite{pareto2018}. In \wechat-AllScene, cold items corresponded to items trained less than 1000 times. Note that, for  a fair comparison with ID-based models, new items with zero training instances were not included in the evaluation. Table~\ref{tab:colditem} shows the results.

As we can see, \precise-UT consistently outperforms all SOTA models, with a $+13\%$ increase from the runner-up model HLLM.
The recommendation efficacy of \sas~and HSTU is significantly compromised due to the absence of textual enhancement, since pure ID models are insufficiently trained for cold items. The decline in text-based LEARN is relatively smaller; however, LEARN fixes token embedding and relies on the model parameters to adapt to collaborative signals, leading to a severe drop when training data is scarce. 
Although HLLM can  leverage the structure of large language models for training and performs optimally in the baseline scheme, it fails to utilize collaborative representation to capture behavioral signals conveyed by item IDs, resulting in slightly inferior performance compared to our model. This observation indicates that models combining ID and textual information can maintain a relatively high level of performance even for long-tail items.

\begin{table*}[htp]
	\centering
	\caption{Scalability Study in terms of Sequence Length, LLM model scale and Transformer Model Scale.\small (Default configuration of \precise~is using $500$ as sequence length, Qwen2-1.5B as LLM model and $6$-layer Transformer. Default configurations are underlined.)} 
    \renewcommand\arraystretch{1.2}
	\begin{tabular}{c|c|c|c|c|c|c|c|c}
		\toprule
		Dataset  &  Direction & Config  &   R@10  & R@30  & R@50 &  N@10  & N@30  & N@50\\
		\hline
		\multirow{9}{*}{\wechat-AllScene} 	&	\multirow{3}{*}{Sequence Length} & 100 & 0.0168
 & 0.0313  & 0.0411 & 0.0114 & 0.0151  & 0.0177  \\ 
	&	& 300 & 0.0186 & 0.0349 & 0.0442 & 0.0130  & 0.0161 & 0.0186  \\
    &	& \underline{500} & 0.0197 & 0.0359 & 0.0460 & 0.0139  & 0.0165 & 0.0185  \\ \cline{2-9}
 & \multirow{3}{*}{LLM Model Scale} & Qwen2-0.5B & 0.0169 & 0.0311  & 0.0405 & 0.0098  & 0.0140  & 0.0173  \\ 
	&	& \underline{Qwen2-1.5B} & 0.0197 & 0.0359 & 0.0460 & 0.0139  & 0.0165 & 0.0185 \\ 
    &	& Qwen2-7B  & 0.0214  & 0.0411 & 0.0547 & 0.0150 & 0.0184 & 0.0217  \\ \cline{2-9}
	&	 \multirow{3}{*}{Transformer Model Scale} & 2 layers & 0.0166 & 0.0305  & 0.0400 & 0.0112  & 0.0152  & 0.0175   \\ 
	&	& \underline{6 layers} & 0.0197 & 0.0359 & 0.0460 & 0.0139  & 0.0165 & 0.0185  \\ 
        &	& 10 layers & 0.0201 & 0.0399 & 0.0535 & 0.0142  & 0.0178 & 0.0198 \\ 
	\bottomrule
	\end{tabular}
    \vspace{-0.2cm}
	\label{tab:scalability}
\end{table*}

\subsection{Ablation Studies of Inputs} As discussed above, \precise-UT outperforms baseline methods based on its utilization of ID and text information. Next, we perform ablation studies of ID and text embedding, as well as the MoE network for combining token embeddings.
In \precise-ID, we remove the LLM module and only use ID for embedding module. Note that \precise-ID is similar to \sas. In \precise-LLM, we remove the ID embedding module and only use LLM encoding for semantic representation. In \precise-Pool, we use average pooling to combine token embeddings of an item without the adaptation of MoE layers. Note that the modified models are based on and compared with \precise-UT model. We report the results using the \wechat-AllScene dataset in  Table~\ref{tab:inputs}.
\begin{table}[htp]
	\centering
	\caption{Performance of Modified Fusion Module. \small (\precise-UT uses the default model design. \precise-ID only uses ID embedding. \precise-LLM only uses LLM token embedding. \precise-Pool uses average pooling instead of MoE network.)  } 
    \renewcommand\arraystretch{1.2}
	\begin{tabular}{c|c|c|c|c}
		\toprule
		 Method &   R@10  & R@30  &  N@10  & N@30  \\
		\hline
		\precise-UT & \textbf{0.0197} & \textbf{0.0359}  & \textbf{0.0139}   & \textbf{0.0165}    \\ 
		 \precise-ID & 0.0106 & 0.0162 & 0.0074 & 0.0093  \\ 
		 		\precise-LLM & 0.0132 & 0.0287 & 0.0088 & 0.0124  \\  
  		\precise-Pool & 0.0186 & 0.0321 & 0.0120 & 0.0157   \\  
		\bottomrule
	\end{tabular}
	\label{tab:inputs}
\end{table}

As depicted in the table, the \precise-UT model outperforms all other variant models. Utilizing ID embeddings alone yields the poorest results  while employing LLM embeddings in isolation performs better. This suggests that LLM possesses an inherent advantage in representing items. Directly employing the pooling  scheme  to join token embeddings 
and concatenate with ID embedding 
results in a significant improvement compared with \precise-ID and \precise-LLM. However, when we integrate token embeddings  using the MoE  model, the performance is further enhanced since the importance of  token embeddings is  determined by the Gating Network  based on collaborative signals.

\begin{table*}[htp]
	\centering
        \renewcommand\arraystretch{1.2}
	\caption{Effectiveness of \precise~Framework in Target Tasks.\small(\precise~adopts both Universal Training and Targeted Training. \precise-UT only uses Universal Training and \precise-TT only uses Targeted Training.)} 
	\begin{tabular}{c|c|c|c|c|c|c|c|c|c}
		\toprule
		Task & Method & \makecell{Universal Training\\Data} &  \makecell{Targeted Training\\Data} &   R@10  & R@30  &  R@50 & N@10  & N@30  & N@50\\
		\hline
		\multirow{3}{*}{Click} 	& \precise	& \wechat-AllScene & \wechat-Click & 0.0158 & 0.0357 & 0.0504 & 0.0104  & 0.0162 & 0.0199   \\ 
		& \precise-UT  &  \wechat-AllScene  & 		-  & 0.0136 & 0.0333 & 0.0492 & 0.0084  & 0.0143 & 0.0182  \\  
  & \precise-TT & - &	\wechat-Click & 0.0109 & 0.0261 & 0.0381 & 0.0069 & 0.0114 & 0.0144 \\  
		\hline
		\multirow{3}{*}{Share} & \precise	&	\wechat-AllScene & \wechat-Share & 0.0055 & 0.0140  & 0.0207 & 0.0031  & 0.0053 & 0.0067  \\ 
	& \precise-UT	&  \wechat-AllScene  & 		-  & 0.0034 & 0.0087 & 0.0125 & 0.0018  & 0.0032 & 0.0040  \\  
& \precise-TT  & - &	\wechat-Share & 0.0018 & 0.0045 & 0.0069 & 0.0010 & 0.0017 & 0.0022 \\  
		\bottomrule
	\end{tabular}
    \vspace{-0.4cm}
	\label{tab:components}
\end{table*}

\subsection{Scalability}
In this section, we attempt to study the scalability of the \precise-UT model in different directions, \ie training data and model parameters. The scale of the dataset in training language models considers the amount of textual sentences. In our experiment, we vary the length of the behavior sequence from $100$ to $500$ as longer sequences bring in more training samples in the Next-item Prediction task.
The scale of parameters comes in twofold in this paper. Firstly, the LLM model that generates the token embedding determines the quality of semantic representation. As reported in many studies, the scale of LLM model parameters largely determines the quality of LLM's output. We evaluate \precise-UT given different scales of LLM models.
Secondly,  the scale of transformer layer parameters is another determining factor in \precise-UT. We report the performance of \precise~ given different number of layers of transformer blocks. The results are reported with the industrial dataset \wechat-AllScene in Table~\ref{tab:scalability}.

As demonstrated in the table, when the sequence length is truncated to $100$ or $300$, there is a corresponding decline in both recall and NDCG performance. This degradation is attributed to the reduction in sequence length, which concurrently decreases the number of training samples, leading to a diminished exposure of user behavior patterns for the model to learn. This suggests that increasing the scale of training data can  further enhance model performance.

In terms of LLM scale, we employed variants of the $Qwen2$ model with different parameter sizes, \ie  0.5B, 1.5B, and 7B. According to the authors' descriptions~\cite{qwen2}, the Qwen2 model is capable of achieving performance improvements as its scale expands. Similarly, we observed a significant performance boost as the scale of the Qwen2 model increases. This further indicates that, on one hand, textual representation plays a crucial role in \precise~and on the other hand, \precise~effectively captures and integrates the enhancements provided by increased model capacity. 

Lastly, we varied the number of layers in the transformer architecture to assess \precise's scalability. It is observable that as we increased the number of layers from 2 to 6 and further to 10, \precise's performance exhibited a progressive increase. This suggests that our model, like the majority of large language models, possesses scalability in terms of model parameters, and we can anticipate achieving superior results by further deepening the model's parameters.
Based on the experimental results above, we conclude that \precise~exhibits multifaceted scalability. This scalability encompasses various aspects, including the volume of data, the selection of the large language models, and the parameter scale of \precise~itself. In future, we anticipate conducting more rigorous and in-depth studies to explore the scalability of \precise~as the scale is expanded across different dimensions.

\subsection{Effectiveness of Framework}
\label{subsubsec:framework}
In this section, we discuss the effectiveness of our designed progressive training framework, which increments from  Universal Training to Targeted Training.  We use \textit{\wechat-Click} and \textit{\wechat-Share} for targeted training for two tasks, \ie clicking and the sharing respectively. 
\wechat-click  collects users' clicking behaviors in a specific scenario and the clicking task is used to test \precise's ability to transform from all-scene click behaviors to target-scene behaviors. 
\wechat-share consists of sharing behaviors from all scenarios and the sharing task is used to test \precise's  transferability in cross-task situations. We compare \precise~with models that only utilize Universal Training (\precise-UT) and  models that only utilize Targeted Training (\precise-TT). Results are shown in Table~\ref{tab:components}.

As illustrated in the table above, we observe that \precise~demonstrates the best performance in both the click and share tasks. This indicates that learning a global representation through Universal Training and subsequently adapting this representation to the target tasks using Targeted Training is an effective approach.  Notably, in the compared methods, directly utilizing the results from \precise-UT for recommendations in either the click or share tasks significantly outperforms the outcomes from directly training \precise-TT in the target scenarios or tasks. This disparity arises because recommendations for target scenarios or tasks often involve  sparse positive samples. Without extensive training with all-scene data, the model is prone to under-fitting, leading to a substantial decline in performance. 
We see that the \precise~framework is effective in terms of utilizing all-scene behaviors to handle data sparsity problems that are commonly observed in real-world recommendation systems.

\section{Online Experiments}
\label{sec:experiments-online}
We apply \precise~in article recommendation services in \wechat. 
In this section, we present the strategies employed for deploying our pre-trained embeddings into online services in detail, along with an analysis of the online performance achieved in recall and ranking.

\begin{table*}[htp]
	\centering
            \renewcommand\arraystretch{1.2}
	\caption{Online Experimental Results in Recall. \small(\#Participants gives the number of users involved in the experiments. Statistically significant improvements are put in bold.)} 
	\begin{tabular}{c|c|c|c|c|c|c|c}
		\toprule
		User Type & Target  &  \makecell{Serving\\Scheme}  &   \#Participants & \#Active User & \#Clicks &  \#Shares  & Reading Time  \\
		\hline
 \multirow{4}{*}{All} 	&	\multirow{2}{*}{Click} & U2I	&	 56 million & \textbf{+0.177\%}  & \textbf{+0.251\%} & +0.983\% &  \textbf{+0.318\%}   \\ 
      &  & U2I2I	&	 10 million & +0.109\%  & \textbf{+0.410\%} & +0.695\%  & +0.361\%  \\  \cline{2-8}
	&	\multirow{2}{*}{Share} & U2I	&	 26 million & +0.054\%  & +0.241\% & \textbf{+2.560\%} & +0.204\%   \\ 
     &   & U2I2I	& 26 million & +0.059\%  & +0.199\% & \textbf{+1.241\%}  & +0.187\%  \\ \hline
 \multirow{4}{*}{Cold-start} 	&	\multirow{2}{*}{Click} & U2I	&	 39 million & \textbf{+0.155\%}  & \textbf{+0.243\%} & -0.219\% &  +0.179\%   \\ 
      &  & U2I2I	&	 30 million & \textbf{+0.182}\%  & +0.077\% & -0.324\%  &-0.074\%  \\  \cline{2-8}
	&	\multirow{2}{*}{Share} & U2I	&	 30 million & +0.027\%  & +0.203\% & +0.482\% & \textbf{+0.268\%}   \\ 
     &   & U2I2I	& 22 million & +0.032\%  & \textbf{+0.348\%} & +0.486\%  & +0.239\%  \\ 
  \bottomrule
	\end{tabular}
	\label{tab:exp-recall}
    \vspace{-0.45cm}
\end{table*}

\subsection{Training and Prediction}

\wechat, being a major platform where users can publish original articles, offers multiple channels for article viewing. These include the Subscription Channel (showing only subscribed accounts' articles), Friend Circle (displaying only friends' shared articles), and Recommendation Box (employing a recommendation system).

For Universal Training, user behaviors from March 2023 to November 2024 across all these scenarios are collected to form the training data. For Targeted Training, two tasks are considered: clicking and sharing. Clicking behaviors in the Recommendation Box are used for the clicking task's Targeted Training, while sharing behaviors from all scenes are used for the sharing task's Targeted Training.

\begin{figure}[h]
\centering
\includegraphics[width=1.0\columnwidth]{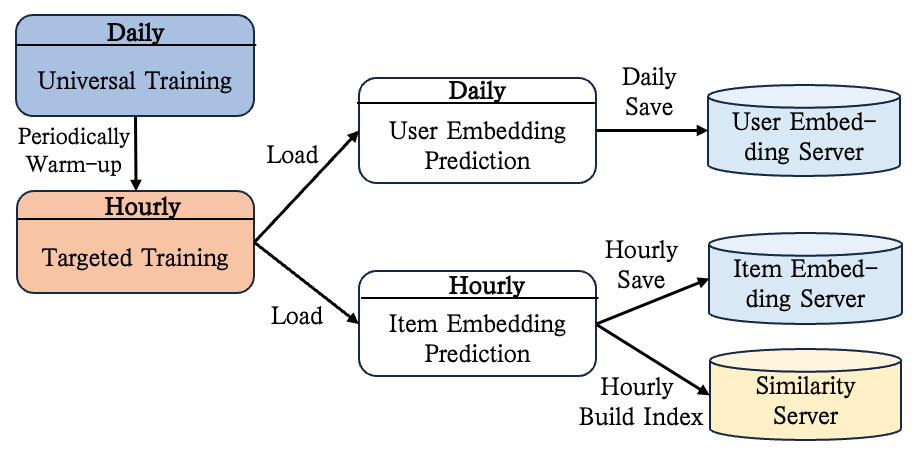}
\caption{The Training and Prediction Scheme  of \precise.}
\label{fig:exp-training}
\vspace{-0.2cm}
\end{figure}

The training process is illustrated in Figure~\ref{fig:exp-training}. The Universal Training model is trained daily. Subsequently, the Targeted Training model loads the parameters from the Universal Training model and incrementally trains for the target tasks. The training strategies introduced in Section~\ref{Sebsec:methodology-discussion}  are utilized.

During the prediction phase, due to the large number of users, full-scale predictions for users are performed daily. The prediction model loads the latest checkpoint from the Targeted Training model each day for user embedding prediction. For items, hourly predictions are conducted to ensure the inclusion of the most recently published items. The embedding results of users and items are saved in the user embedding server and item embedding server respectively. These servers enable downstream applications to quickly retrieve embeddings given query IDs. Additionally, a similarity server is used to construct an index of all item embeddings, facilitating the rapid retrieval of the item list with the highest cosine similarity for any query embedding.

\subsection{Online Serving in Recall}
\begin{figure}[h]
\centering
\includegraphics[width=1.02\columnwidth]{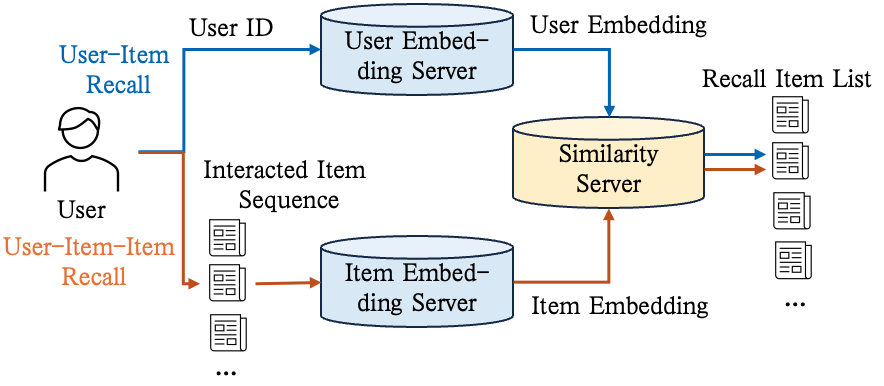}

\caption{Serving Scheme  of \precise~in Recall.}
\label{fig:exp-recall}
\vspace{-0.3cm}
\end{figure}

Leveraging the embedding server and similarity server~\cite{faiss2024}, \precise's serving in the recall phase has two paradigms: User - Item (U2I) recall and User - Item - Item (U2I2I) recall. Figure~\ref{fig:exp-recall} shows the details.
In U2I recall, the user embedding is retrieved from the user embedding server using the user's ID as the query key. Then, it is input to the similarity server to obtain the item IDs with the least cosine distance in embedding.
For U2I2I recall, users' interacted item sequences are first retrieved based on the target task. That is, for the clicking task, users' recently clicked items are retrieved, and for the sharing task, users' recently shared items are retrieved. Then, these items' pre-trained embeddings are retrieved from the item embedding server. These item embeddings are input to the similarity server to obtain the most similar items as recall results.


Both U2I and U2I2I recall paradigms are deployed in the Article Recommendation task. The pre-trained embedding-based recall is augmented for two types of user populations: the entire user base and cold-start users. The baseline group uses popular recall methods like DSSM~\cite{dssm2013} and GNN~\cite{gnn2023}. The experimental groups add the corresponding U2I and U2I2I methods as additional recall methods.

We have deployed both U2I and U2I2I recall paradigms in the Article Recommendation task.
We augmented the pre-trained embedding-based recall for two types of  user populations: the entire user base and cold-start users. Baseline group uses popular recall methods \ie DSSM~\cite{dssm2013} and GNN~\cite{gnn2023} trained in a streaming scheme. Experimental groups add the corresponding U2I and U2I2I recall methods based on pre-trained embeddings.
For the sake of privacy, we only report relative improvements in this paper. 
Table~\ref{tab:exp-recall} show the results.  Figures in bold indicate statistical significance with a P-value below $0.05$ in the significance test.

As we can see, we take notes of four important online metrics, namely total number of active users, total number of clicks, total number of shares and total reading time among all participated users.
For the entire user base, we conducted a total of four experiments, utilizing embeddings trained on clicking and sharing data  for both U2I and U2I2I paradigms. For the clicking task, there is a significant enhancement in recall performance by directly employing the U2I approach. All click-related metrics show substantial improvements. Specifically, the number of active users increased by $0.177\%$, and the reading duration improved by $0.318\%$. When we deploy the U2I2I method, there is a notable increase in the number of clicks ($+0.410\%$). It is worth noting that although the increase in reading duration with the U2I2I method is greater than with the U2I method ($+0.361\%$ compared to $+0.318\%$), the improvement with U2I2I is not statistically significant. This discrepancy arises because the two experiments were conducted on different user populations, and the significance test results vary with the activity level of these populations.
Next, we examine the effects of using embeddings trained on sharing data in the recall process. It can be observed that employing the U2I recall method significantly increased the number of online sharing behaviors by $2.560\%$. The U2I2I method also significantly increased the number of shares by $1.241\%$. The relatively smaller increase with U2I2I is due to its reliance on users having historical positive example sequences, which is less favorable for cold-start users in the specific scenario. However, for U2I recall, since we train the sequence using data from all scenarios, a user's behavior in other scenarios can  assist in tasks for the target scenario, thereby achieving better results.

\begin{table*}[htp]
	\centering
        \renewcommand\arraystretch{1.2}
	\caption{Online Experimental Results in Ranking.\small(\#Participants gives the number of users involved in the experiments. Statistically significant improvements are put in bold.)} 
	\begin{tabular}{c|c|c|c|c|c|c|c}
		\toprule
		 Method & \makecell{Universal Training\\Data} &  \makecell{Targeted Training\\Data} &    \#Participants & \#Active User & \#Clicks &  \#Shares  & Reading Time  \\
		\hline
		 \precise	& All Scene Click & Target Scene Click &  180 million & \textbf{+0.174\%} & \textbf{+1.961\%}   & +1.433\% & \textbf{+0.884\%} \\ 
		 \precise-UT  &  All Scene Click  & 		-  &  180 million & +0.079\% & \textbf{+1.649\%} &  +1.961\% &  \textbf{+1.287\%} \\
		\bottomrule
	\end{tabular}
	\label{tab:exp-ranking}
    \vspace{-0.4cm}
\end{table*}

Cold-start users or inactive users are defined as those individuals who have read articles in the Recommendation Box scenario less than a specific number of times within recent days. These users constitute approximately $90\%$ of the total user population. For such cold-start users, we also implement four types of recalls as supplementary measures through the U2I and U2I2I recall strategies.
It can be noted that when enhancing the recall process with pre-trained click embeddings, there is a notable increase in the number of active users. Specifically, for U2I and U2I2I, the increases are $+0.155\%$ and $+0.182\%$ respectively. This clearly demonstrates that pre-trained click embeddings have a beneficial effect in converting non-clickers into clickers. Additionally, using click-based U2I for recall also leads to a significant rise in the total number of clicks ($+0.243\%$).
Nevertheless, there is a marginal decrease in the number of shares, suggesting a certain seesaw phenomenon. That is, it is challenging to simultaneously enhance both clicks and shares for cold users. When employing sharing embedding as a recall method, we observe that it is difficult to boost sharing behaviors among inactive users. This is because sharing demands a deeper level of user engagement compared to clicking. Despite this, we find that the pre-trained sharing embedding has a positive influence on clicks and reading time. This implies that modeling sharing behavior across all scenarios can capture certain  user interests, thereby enhancing the dwell time for cold-start users.

\subsection{Online Serving in Ranking}
\begin{figure}[h]
\centering
\includegraphics[width=0.85\columnwidth]{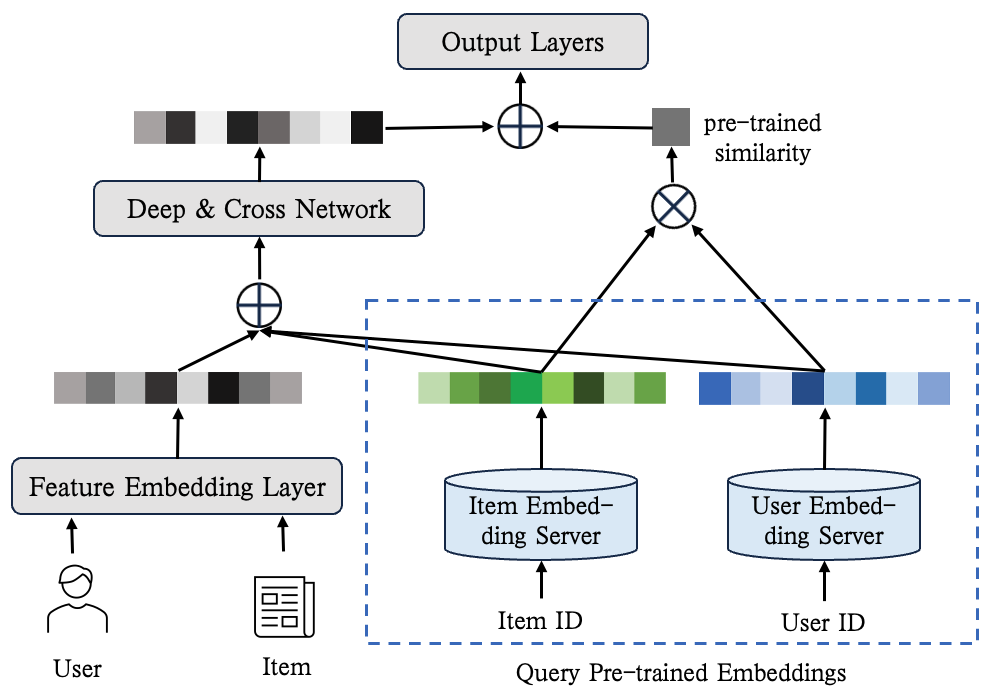}
\caption{Serving Scheme  of \precise~in Ranking.}
\label{fig:exp-ranking}
\vspace{-0.55cm}
\end{figure}

In the online ranking phase, we adopt pre-trained user and item embeddings as features to enhance the performance of the ranking model. In a typical recommendation system's ranking model, the discrete features of users and items are first embedded through an embedding layer and then input into a Deep \& Cross Network (DCN) architecture for feature interaction~\cite{dcn2021}. Eventually, they are fed into the output layers to obtain the ranking results. To incorporate the pre-trained embeddings, 
we perform two operations as shown in Figure~\ref{fig:exp-ranking}: 1) we concatenate the pre-trained embeddings with the output of the ranking model's embedding layer. The concatenated embedding is further fed into the DCN network for feature crossing. 2) we compute the dot product of the user and item embeddings and concatenate the resulting score with the output embedding from the ranking model's DCN module. Through these operations, the ranking model can treat the pre-trained embeddings as feature inputs and utilize the pre-trained scores to evaluate the similarity between users and items.


We integrate the pre-trained embeddings into a ranking model that evaluates the clicking probability of user-item pairs. Since the ranking model focuses on predicting clicks, we conduct experiments with two \textit{click} embeddings. The first embedding is generated by \precise-UT with Universal Training from all-scene clicks and without Targeted Training. The second embedding is generated by \precise~with Universal Training on all-scene clicks and Targeted Training on target-scene clicks. The experimental results (shown in Table~\ref{tab:exp-ranking}) indicate that incorporating the embeddings from either model as features into the ranking model leads to a certain degree of performance improvement. Both models significantly increase the number of clicks and reading duration, suggesting that the pre-trained models can effectively provide information gain to the ranking model. When comparing the two models, the \precise~model shows a superior advantage in the number of active users ($+0.174\%$) and the number of clicks ($+1.961\%$)  because it is trained with the click behaviors of the target scenario. This result is consistent with our previous conclusions based on offline experiments in Section~\ref{Subsec:methodology-framework}, demonstrating the effectiveness of our framework in improving the performance of the target task in the target scenario. The other model, \precise-UT, exhibits a relatively less obvious enhancement in the number of active users and clicks. However, \precise-UT shows a more advantageous improvement in reading time ($+1.287\%$ compared to $+0.884\%$). This may be attributed to the model's introduction of a broader range of user interests from other scenarios.

\section{Conclusion}
\label{Sec:conclusion}
In this paper, we propose \precise, a sequence recommendation pre-training framework. \precise~adopts an MoE structure to effectively combine token embeddings from LLMs. The combined embeddings are further concatenated with ID embedding for integrating  semantic information with collaborative behavior.
Subsequently, through a Universal Training plus Targeted Training paradigm that transitions from a holistic scenario to a target scenario or task, we model the comprehensive behavior of users and rapidly adapt to target interests. 
We have achieved superior performance compared to state-of-the-art solutions on both public and industrial datasets. Most importantly, our model framework has been fully deployed online and has yielded significant benefits. 
Future work includes research methods for faster convergence of Targeted Training models. 
Additionally, the application of pre-trained embeddings in industrial recommendation systems is worth exploring. 
This will help to further improve the performance and applicability of \precise~in real-world scenarios.

\newcommand{\RefFile}{main}
\bibliographystyle{IEEEtran}
\bibliography{\RefFile}

\end{document}